# Field-Induced Local Excitations Causing Zero-Magnetization Plateaus in Antiferromagnets of Antiferromagnetic Spin Dimers Under Magnetic Field

Myung-Hwan Whangbo[1,*], Hyun-Joo Koo[2], Nikita V. Astakhov,[3,4] Peter S. Berdonosov,[4] Olga S. Volkova[5,*]

[1] Department of Chemistry, North Carolina State University, Raleigh, NC 27695-8204, USA

[2] Department of Chemistry, Research Institute for Basic Sciences, Kyung Hee University, Seoul 02447, Republic of Korea

[3]Faculty of Materials Sciences, Lomonosov Moscow State University, Moscow 119991, Russia

[4]Faculty of Chemistry, Lomonosov Moscow State University, Moscow 119991, Russia

[5]Faculty of Physics, Lomonosov Moscow State University, Moscow 119991, Russia

mike_whangbo@ncsu.edu

os.volkova@yahoo.com

## Abstract

The zero-magnetization plateau refers to the phenomenon that the magnetization of an antiferromagnet under magnetic field remains at zero when the field increases from 0 to a certain critical value $\mu_0 H_c$. Certain antiferromagnets composed of antiferromagnetic (AFM) spin dimers exhibit a zero-magnetization plateau despite that the single-ion anisotropy of their magnetic ions is negligible. To investigate the cause for this finding, we analyzed how a magnetic field affects the energy spectrum of an AFM chain composed of AFM spin dimers made up of two $S = 1/2$ ions under the supposition that each spin dimer counteracts external field, according to Le Chatelier's principle, to introduce a small amount of spin into the ground state ($S = 0$) by mixing the excited state ($S = 1$). First, we show that this concept of field-induced local excitations explains several puzzling observations concerning the zero-magnetization plateaus of $TlCuCl_3$ and $KCuCl_3$, which are antiferromagnets made up of molecular anions $Cu_2Cl_6^{2-}$ (i.e., spin dimers of $S = 1/2$ ions $Cu^{2+}$). Then, we analyze the magnetic entropy associated with the zero-magnetization plateaus occurring in the field region between 0 and $\mu_0 H_c$ to find that this field region is divided into two subregions of different magnetic entropy; the $0 – \mu_0 H_m$ region where two different magnetic species coexist, the distribution of which in the spin lattice generates the magnetic entropy described largely by the binomial coefficients, and the $\mu_0 H_m – \mu_0 H_c$ region where there exists only one magnetic species, leading to zero magnetic entropy. We carried out specific heat measurements for $KCuCl_3$ as a function magnetic field between 0 – 9 T at 2 K and verified the predictions of our analysis concerning the field-dependence of specific heat for the two-phase region of a zero-magnetization plateau.

## 1. Introduction

The magnetization ($M$) vs. magnetic field ($\mu_0 H$) curve of an antiferromagnet may remain constant in certain region(s) of the field typically at an integer fraction of its saturation magnetization (e.g., $M/M_{sat} = 1/3$). The occurrence of a nonzero magnetization plateau in such an antiferromagnet is well explained by the supposition[1] that it counteracts the field by partitioning its AFM spin lattice into ferrimagnetic fragments of nonzero spin $S$, hence absorbing Zeeman energy, $E_Z = \mu_0 \mu_B g \vec{S} \cdot \vec{H}$, in accordance with Le Chatelier's principle. A theoretical basis for this supposition has recently been provided by analyzing how the external magnetic field affects the magnetic entropy of an antiferromagnet exhibiting a nonzero-magnetization plateau.[2] Certain antiferromagnets exhibit a zero-magnetization plateau, namely, their magnetization remains at $M/M_{sat} = 0$ when the field increases from 0 to a critical value $\mu_0 H_c$. There are two types of zero-magnetization plateaus (hereafter, type I and type II) to consider.

A zero-magnetization plateau of type I occurs in an antiferromagnet with substantial single ion anisotropy D.[3,4] Below its Néel temperature, $T_N$, the spin moments of such an antiferromagnet are ordered along the preferred direction dictated by D. Under applied field, this preferred spin orientation remains unaffected until the field reaches a particular value, $\mu_0 H_{sf}$, at which point a spin-flop occurs, i.e., the spins flop to the orientation perpendicular to the field and then begin to tilt toward the field direction as the field increases beyond $\mu_0 H_{sf}$ (**Fig. 1a**). (Thus, $\mu_0 H_{sf} = \mu_0 H_c$.) This results in a zero-magnetization plateau between 0 and $\mu_0 H_{sf}$ in the magnetization curve (**Fig. 1b**).[1,5] When the field is perpendicular to the preferred spin orientation (**Fig. 1c**), the field and the spin orientation have the relative arrangement identical to the one found above the spin-flop point (**Fig. 1a**), so a zero-magnetization plateau does not occur.[1,5]

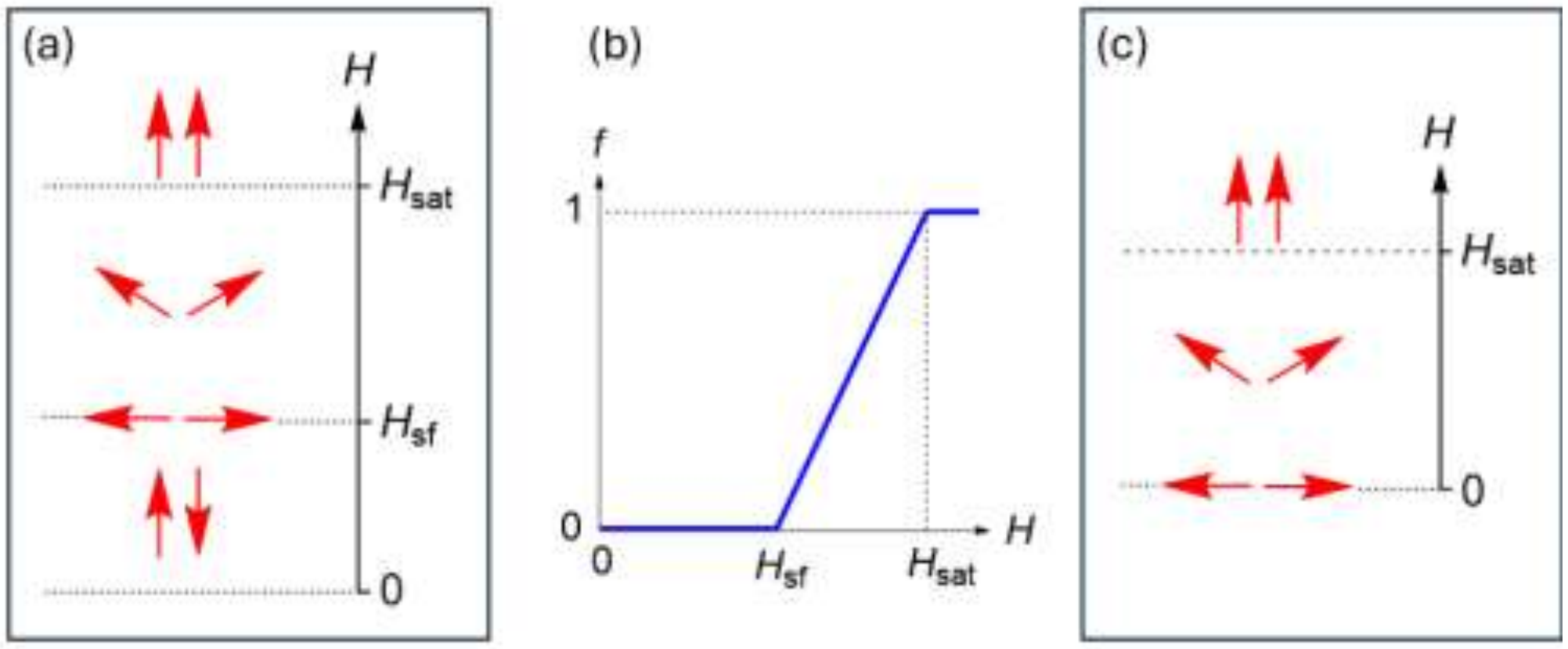

**Fig. 1**. Field-dependence of the spin orientations and the magnetization for an antiferromagnet with substantial single-ion magnetic anisotropy below $T_N$: (a) Change in the orientation of the ordered spin moments as the magnetic field increases along the direction of the ordered moment. Here the arrows represent the moments of the two magnetic sublattices of the antiferromagnet. (b) Schematic representation of the expected magnetization curve with zero-magnetization between 0 and $\mu_0 H_{sf}$. (c) Change in the orientation of the ordered spin moments as the magnetic field increases as the magnetic field increases along the direction perpendicular to the ordered moment.

A zero-magnetization plateau of type II is found for an antiferromagnet of antiferromagnetically coupled AFM spin dimers made up of magnetic ions with negligible single-ion anisotropy, as found for molecular magnets $KCuCl_3$ and $TlCuCl_3$,[1,6] which consist of $Cu_2Cl_6^{2-}$ anions in which two magnetic ions $Cu^{2+}$ ($d^9$, $S = 1/2$) are antiferromagnetically coupled (see below). For the convenience of our discussion, we will use the convention that AFM and ferromagnetic (FM) spin exchanges are represented by positive and negative J values, respectively. In addition, we use the term “a magnetic bond” to mean an AFM (FM) spin exchange path whose two magnetic ions are antiferromagnetically (ferromagnetically) coupled, “a magnetic antibond” to mean an AFM (FM) spin exchange path whose two magnetic ions are ferromagnetically (antiferromagnetically) coupled, and "breaking a magnetic bond" to mean the conversion from a magnetic bond to a magnetic antibond.

For simplicity, let us consider an AFM chain of antiferromagnetically-coupled AFM dimers (**Fig. 2a**), in which the intradimer spin exchange J is stronger than the interdimer spin exchange J′ (i.e., J > J′). It is convenient to separate the AFM dimers into two sets by choosing every second AFM dimer, which is indicated by black and purple colors in **Fig. 2b**. The AFM dimers begin to break their J′ bonds under magnetic field, and the number of broken J′ bonds increases with field. This is achieved by spin-flipping the AFM dimers, one at a time (indicated by red spheres in **Fig. 2c**), in one set of the AFM dimers. This continues until all the AFM dimers are spin-flipped in one set of the AFM dimers (**Fig. 2d**) at a certain field $\mu_0 H_\mathrm{m}$. As the field increases beyond $\mu_0 H_\mathrm{m}$ and becomes strong enough to break the intradimer bond J at a certain field $\mu_0 H_\mathrm{c}$, the intradimer bond J begins to break one at a time (**Fig. 2e**) with increasing field beyond $\mu_0 H_\mathrm{c}$, and eventually all intradimer bonds are broken at a certain field $\mu_0 H_{\mathrm{c}'}$ (**Fig. 2f**). A schematic magnetization vs. magnetic field curve describing the AFM chain of AFM dimers is presented in **Fig. 2g**. Since $S$ = 0 for each AFM dimers, the magnetization remains zero until all the interdimer bonds are broken at $\mu_0 H_\mathrm{m}$. Furthermore, the magnetization remains zero beyond $\mu_0 H_\mathrm{m}$ up to $\mu_0 H_\mathrm{c}$, and each AFM dimer starts to become fully polarized beyond $\mu_0 H_\mathrm{c}$. That is, the width of the zero-magnetization plateau is solely determined by the strength of the intradimer bond J.

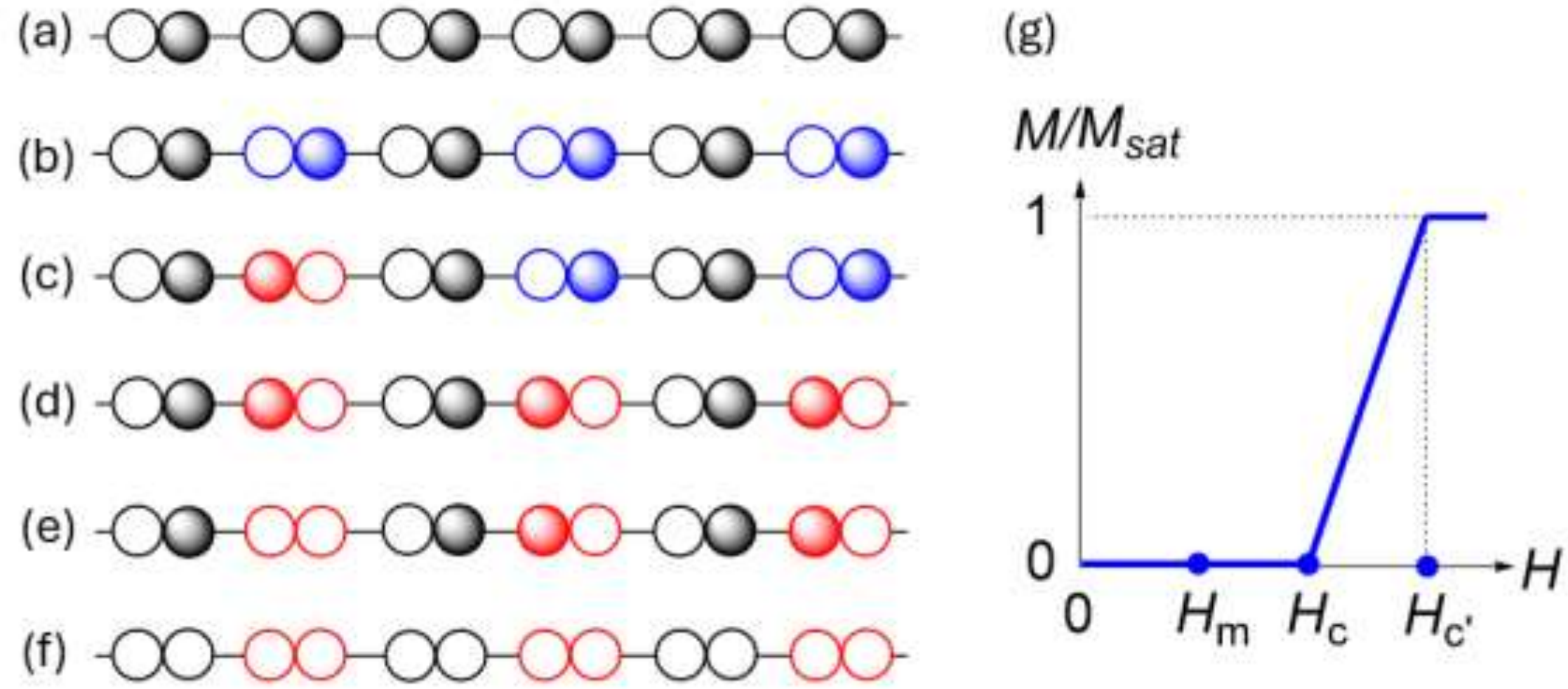

**Fig. 2**. (a) AFM chain of antiferromagnetically coupled AFM dimers, for which the intra- and inter-trimer spin exchanges (J and J′, respectively) are both AFM with J > J′. Here the unshaded and shaded spheres represent the up- and down-spins. (b) Separation of the AFM chain of AFM dimers into two sets, which are distinguished by black and purple colors. (c) AFM chain of AFM dimers with one spin-flipped AFM dimer (represented by red spheres) in the purple set of AFM dimers. Each spin-flipped AFM dimer breaks two interdimer bonds J′. (d) Chain of AFM dimers of ferromagnetically coupled AFM dimers, which is generated by spin-flipping all AFM dimers in the purple set. (e) Chain of ferromagnetically coupled AFM containing a fully polarized (i.e., FM) dimer. (f) Chain of ferromagnetically coupled FM dimers, which is generated when all AFM dimers become fully polarized. (g) The magnetization curve expected for the AFM chain of antiferromagnetically coupled AFM dimers.

The two types of zero magnetization plateaus discussed above raise a conceptual issue. Suppose that an external magnetic field is applied along the preferred moment direction of an antiferromagnet that exhibits type I zero magnetization plateau (**Fig. 1a**). Then $E_Z = 0$ in the 0 – $\mu_0 H_{sf}$ region because the Zeeman energy of the up-spin moment cancels that of the down-spin moment. This leads one to ask how an antiferromagnet of type I magnetization plateau accumulates enough energy in the 0 – $\mu_0 H_{sf}$ region to spin-flop at $\mu_0 H_{sf}$. (When the field is greater than $\mu_0 H_{sf}$, the moments of both spin sublattices tilt toward the field direction leading to a nonzero $E_Z$. As the field increases, the antiferromagnet counteracts more strongly by tilting the moments more hence increasing the associated Zeeman energy. The same holds true when the field is applied along the direction perpendicular to the preferred orientation of the moments.) The same question arises for an antiferromagnet of type II magnetization plateau because AFM spin dimers cannot absorb Zeeman energy since $S$ = 0 for the spin dimers. Accordingly, we investigate how an antiferromagnet, exhibiting either type I or type II zero magnetization plateau, absorbs Zeeman energy, resulting in either spin-flop transitions or the breaking of interdimer bonds, to find that the emergence of a zero-magnetization plateau is explained by the supposition that applied field induces local excitations in spin dimers hence mixing the excited state of S ≠ 0 into the ground state of $S$ = 0. This concept of field-induced local excitations was put forward to explain the apparently puzzling magnetic properties of several antiferromagnets.[2,7-9]

Another conceptual issue is the magnetic entropy associated with type II zero magnetization plateaus. The field region between 0 and $\mu_0 H_m$ (**Fig. 2c**, **2d** and **2g**) describing a type II zero-magnetization plateau is a two-phase region, that is, it consists of AFM dimers with broken interdimer bonds as well as those with no broken interdimer bonds. We analyze how the AFM dimers with broken interdimer bonds are distributed in the AFM spin lattice of AFM dimers and how this distribution governs the field-dependence of configurational magnetic entropy. The predictions of this analysis will be verified by measuring how the specific heat of $KCuCl_3$ depends on magnetic field.

Our work is organized as follows: **Section 2** briefly describes our experimental and our methods of theoretical analyses. In **Section 3**, we put forward the supposition that field-induced local excitations are the driving force for zero-magnetization plateaus and discussed how this hypothesis allows one to explain the anomalous observations on the zero-magnetization plateaus of two molecular solids $KCuCl_3$ and $TlCuCl_3$ as well as the anisotropy in the zero-magnetization plateau of $KCuCl_3$. Then, we described the magnetic entropy associated with type II zero-magnetization plateaus and how it can be experimentally detected by performing specific heat measurements. This point was experimentally verified by performing specific heat measurements for $KCuCl_3$. After discussing the need for further studies on field-dependent magnetic entropy in **Section 4**, we summarize our conclusions in **Section 5**.

## 2. Experimental and methods

Crystalline powder samples of $KCuCl_3$ were prepared by hydrothermal technique. A 25-mL PTFE inlay in stainless steel autoclave was loaded with 1.7060 g of $CuCl_2$ $2H_2O$ (analytical grade, Reachem, Russia), KCl (pure, Reachem, Russia), and 3 mL of concentrated HCl (pure, SIGMATEK LLC, Russia) and heated at 120°C for 14 hours. Brown powder samples with needle-shaped crystals were obtained. The resulting samples were analyzed on a Stoe Stadi-P diffractometer (Stoe&Cie GmbH, Germany) using CuKα1 radiation and an IP detector. The powder diffraction pattern was fully indexed using WinXPOW software (Stoe&Cie GmbH) in a monoclinic unit cell in the space group $P2_1/c$ with the cell parameters a = 4.0315(10) Å, b = 13.7994(24) Å, c = 8.7420(15) Å, β = 97.167(12)°, cell volume 482.53(23) $Å^3$ with a quality factor F(30) = 76.6. These data are in good agreement with those previously reported for $KCuCl_3$.[10] The powder samples were ground in an agate mortar and pressed into a pellet for the specific heat measurements by quasi-adiabatic calorimeter option of Physical Property Measurement System PPMS-9T "Quantum Design".

In our analysis of the driving force for the zero-magnetization plateaus of $KCuCl_3$ and $TlCuCl_3$, it is necessary to know the spin exchanges describing their spin lattices. Matsumoto et al.[11] extracted the "experimental" values for these spin exchanges by analyzing the magnetic excitation energies of $KCuCl_3$ and $TlCuCl_3$ reported by Cavadini et al.,[12,13] who carried out inelastic neutron scattering measurements on single crystal samples. The theoretical values for the spin exchanges, evaluated by performing the energy-mapping analysis[1] based on DFT calculations, were already reported.[1] We employed these two sets of the spin exchanges for our study.

## 3. Results

### 3.1. Field-induced local excitations as the driving force for zero-magnetization plateaus

To find how an antiferromagnet made up of AFM dimers can counteract the field, we consider that each spin dimer unit is made up of two $S$ = 1/2 magnetic ions (**Fig. 3**) in which the two spin sites are antiferromagnetically coupled. Then, the ground and excited states of each spin dimer are described by the singlet ($S$ = 0) and triplet ($S$ = 1) states, respectively. A minute nonzero spin moment, $\delta S$, can be introduced into the spin dimer by the occupation of its triplet state, which is determined by the Boltzmann factor $\exp(-\Delta/k_B T)$ so that

$$\delta S \propto \exp(-\Delta/k_B T), \quad (1)$$

where $\Delta$ is the energy gap between the singlet and triplet states when $\mu_0 H = 0$. (For an isolated spin dimer, $\Delta$ is equal to the intradimer spin exchange J.) Then, the dimer can have nonzero Zeeman energy

$$\delta E_Z = \mu_0 \mu_B g H \delta S. \qquad (2)$$

When this energy becomes large enough with increasing $\mu_0 H$, it can be used to break interdimer bonds.

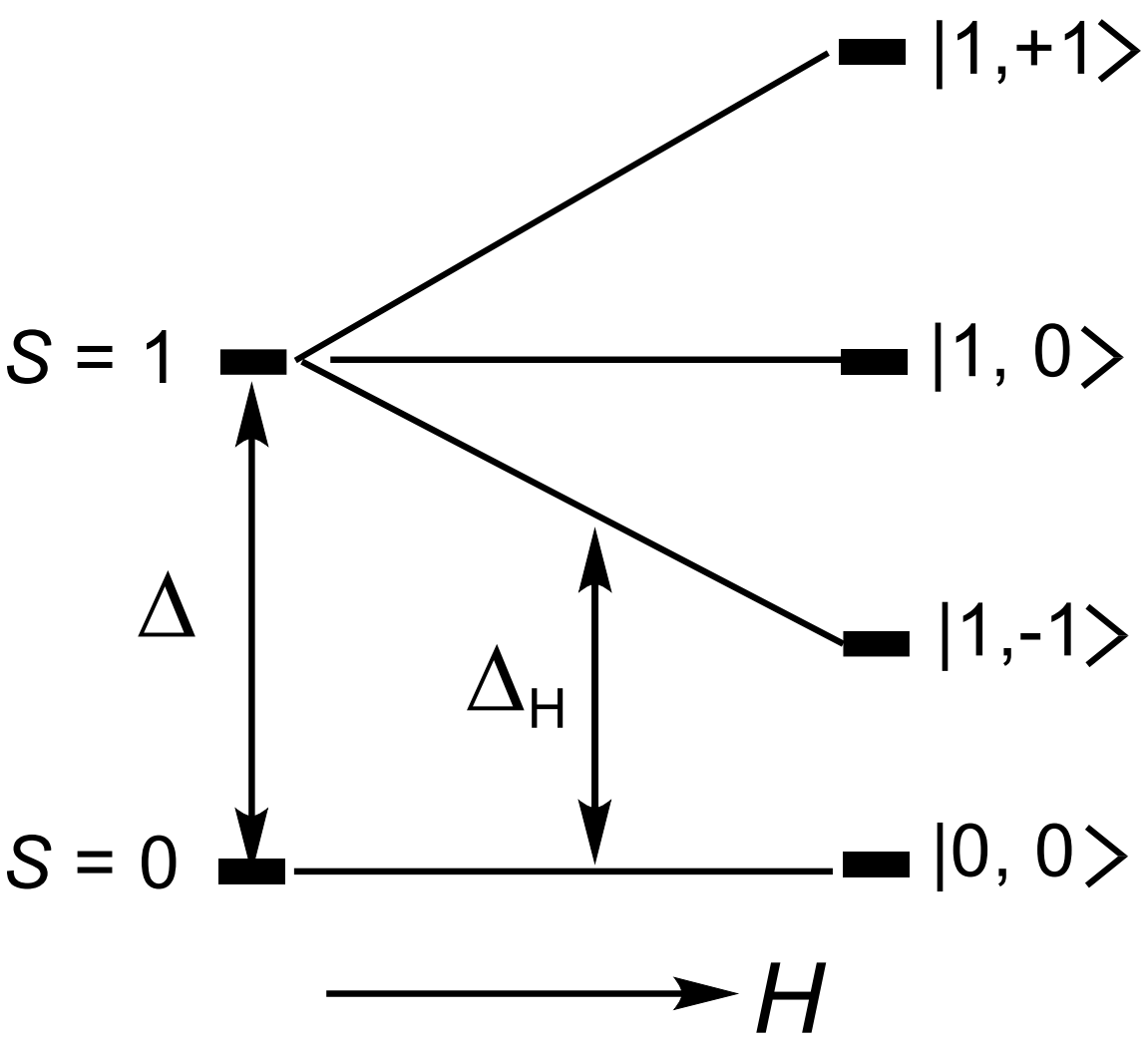

**Fig. 3**. Singlet and triplet states of an isolated spin dimer made up of two antiferromagnetically coupled $S = 1/2$ ions.

Under magnetic field, the triplet state $|S, S_z\rangle$ ($S = 1$, $S_z = 1, 0, -1$) is split such that the substate $|1, -1\rangle$ of the triplet state is lowered in energy linearly with magnetic field, thereby reducing the energy gap $\Delta_H$ between the singlet ground state $|0, 0\rangle$ and the triplet excited state $|1, -1\rangle$. $\Delta_H$ can be written as $\Delta_H = \Delta(1 - H/H_d)$, where $\mu_0 H_d$ is the field at which $\Delta_H$ becomes zero. Then, at $H > 0$, the Boltzmann factor $\exp(-\Delta_H/k_B T)$ for the state $|1, -1\rangle$ is rewritten as

$$\exp[(-\Delta/k_B T)(1 - H/H_d)]. \qquad (3)$$

Since the magnetization measurements are carried out at a very low temperature (i.e., typically at 1 – 4 K), a small difference in Δ can make a huge difference in the Boltzmann factor. (It should be noted that, if spin-orbit coupling is not neglected, the singlet and triplet states mix, though very weakly, such that the singlet state has a small amount of triplet character.[3,4]) For an antiferromagnet made up of general AFM spin dimers, the low-lying excited state of the AFM unit with nonzero spin moment should be responsible for the occurrence of its zero-magnetization plateau.

As discussed above, the most likely driving force enabling the formation of zero-magnetization plateaus is the mixing of the excited state into the ground state as dictated by the Boltzmann factor. This concept of field-induced local magnetic excitations was employed in explaining why the spin moments of $Fe^{3+}$ ($S$ = 5/2) ions of several antiferromagnets in their long-range ordered AFM states are much lower than 5 $\mu_B$ (i.e., 1.56 – 4.48 $\mu_B$)[7-9] and why a smaller magnetic specific heat $C_m(T)$ is observed for γ-$Mn_3(PO_4)_2$ when measured under high magnetic field.[2] The Boltzmann factor, $\exp(-\Delta/k_BT)$, decreases exponentially with the energy gap Δ. Given two antiferromagnets with different energy gaps, therefore, the Boltzmann factor would be much greater for the antiferromagnet with smaller energy gap. We will examine this issue in the next section.

The above analysis leads to an interesting conceptual question. According to Eq. (1) and (3), the value of $\delta S$ is nonzero and will increase with increasing the external magnetic field. Then, strictly speaking, the “zero”-magnetization plateau should mean a magnetization plateau with the magnetization very close to zero. Indeed, the zero-magnetization curves of $KCuCl_3$ reported by Tanaka et al.[6] begin to increase their slopes from zero almost exponentially when the field increases beyond 10 T. This finding in turn supports our proposal that the driving force for the zero magnetization is the mixing of the triplet excited state into the singlet ground state by the Boltzmann factor.

### 3.2. Zero-magnetization plateaus of $ACuCl_3$ (A = Tl, K)

In this section we show that the supposition put forward in the previous section leads one to explain several seemingly puzzling observations concerning the zero-magnetization plateaus of $ACuCl_3$ (A = Tl, K).

### 3.2.A. Crystal structure

The $Cu_2Cl_6^{2-}$ anions in $KCuCl_3$ [10] and $TlCuCl_3$ [11] are made up of two edge-sharing $CuCl_4$ square planar units, each with a $Cu^{2+}$ ($S$ = 1/2) ion (**Fig. 4a**). In $KCuCl_3$ and $TlCuCl_3$, the intradimer spin exchange J is AFM so that the ground and excited states of each $Cu_2Cl_6^{2-}$ anion are the singlet ($S$ = 0) and the triplet ($S$ = 1) states, respectively. The cleavage planes of $KCuCl_3$ and $TlCuCl_3$ crystals are the (102) planes (**Fig. 4b**).[5] The planes of the $Cu_2Cl_6^{2-}$ ions are inclined to the (102) plane by ~20° (**Fig. 4b**) so that the cleavage plane is approximately parallel to the planes of the $Cu_2Cl_6^{2-}$ ions. We represent the fields parallel and perpendicular to the cleavage plane as the $H$||cp and $H$⊥cp fields, respectively. In the magnetization measurements for $KCuCl_3$ and $TlCuCl_3$ performed at 1.7 K,[5] the zero-magnetization plateaus are found to be much narrower for $TlCuCl_3$ than for $KCuCl_3$, namely, ~6 vs ~22 T for the $H$||cp field, and ~6 vs ~20 T for the $H$⊥cp field (**Fig. 4c**). Each $CuCl_4$ unit of the $Cu_2Cl_6^{2-}$ ion can be approximated by an ideal $CuCl_4$ square plane with the four-fold rotational axis as the local z-axis. If the directions parallel and perpendicular to the z-axis are represented by the ||z and ⊥z directions, respectively, then the ||cp and ⊥cp directions are approximately the ⊥z and ||z directions, respectively (**Fig. 4b**).

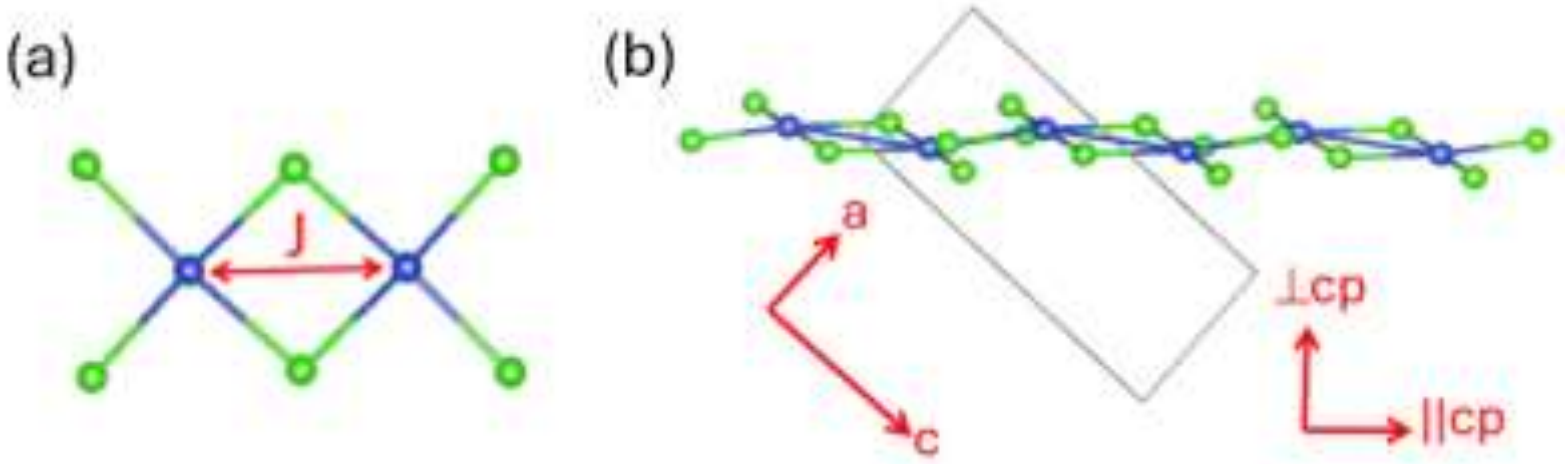

(c) **Width of 0-Magnetization Plateau (T)**

| | ***H*\|\|cp** | ***H*⊥cp** |
|---|---|---|
| **$KCuCl_3$** | **~22** | **~20** |
| **$TlCuCl_3$** | **~6** | **~6** |

**Fig. 4**. (a) $Cu_2Cl_6^{2-}$ anion made up of two $CuCl_4$ square planar units each containing $Cu^{2+}$ ($d^9$, $S$ = 1/2) ion. (b) Projection view of one layer of $ACuCl_3$ (A = K, Tl) parallel to the (102) plane, which is the cleavage plane. Each $Cu_2Cl_6^{2-}$ anion is inclined to the (102) plane by ~20°. (c) The widths of the zero-magnetization plateaus of $KCuCl_3$ and $TlCuCl_3$ single crystals found from the magnetization measurements with field applied along the directions parallel and perpendicular to the cleavage plane.

### 3.2.B. Intradimer spin exchanges governing the width of zero-magnetization plateaus

For $KCuCl_3$ and $TlCuCl_3$, the widths of their zero-magnetization plateaus measured with the *H*||cp and *H*⊥cp fields are similar, indicating that measurements with the *H*⊥z and *H*||z fields should also yield comparable results (**Fig. 4c**). Consequently, the zero-magnetization plateaus of $KCuCl_3$ and $TlCuCl_3$ are not type I but type II.[1,5] (We note that the spin orientations of the $Cu^{2+}$ ions observed for $KCuCl_3$ [14] and $TlCuCl_3$ [15] from neutron diffraction measurements are approximately along the ⊥z directions. The reason for this observation was discussed in **Section S1** with **Fig. S1** in the supporting information.) It should be recalled that the width of the zero-

magnetization plateau is not determined by $\mu_0 H_m$ but by $\mu_0 H_c$, when the interdimer spin bond is weaker than the intradimer spin bond (**Fig. 2g**).

The values of the intra- and interdimer magnetic bonds of $KCuCl_3$ and $TlCuCl_3$, determined by the energy-mapping analysis based on DFT+U calculations,[1,4] are summarized in **Fig. 5**. The $Cu_2Cl_6^{2-}$ spin dimers of $KCuCl_3$ and $TlCuCl_3$ form layers parallel to the bd-plane (i.e., the cleavage plane, **Fig. 5a**), and these layers are stacked along the a-direction (**Fig. 5b**). In each layer, every spin dimer is surrounded by six spin dimers making two $J_2$ and two $J_3$ interdimer bonds (**Fig. 5a**). Along the stacking direction, each spin dimer makes two $J_{a'}$ interdimer bonds. As shown in **Fig. 5c**, all strong interdimer bonds $J_2$, $J_3$ and $J_{a'}$ are weaker than the intradimer bond $J_1$ in both $TlCuCl_3$ and $KCuCl_3$, and the intradimer bond $J_1$ is weaker for $TlCuCl_3$ than for $KCuCl_3$.

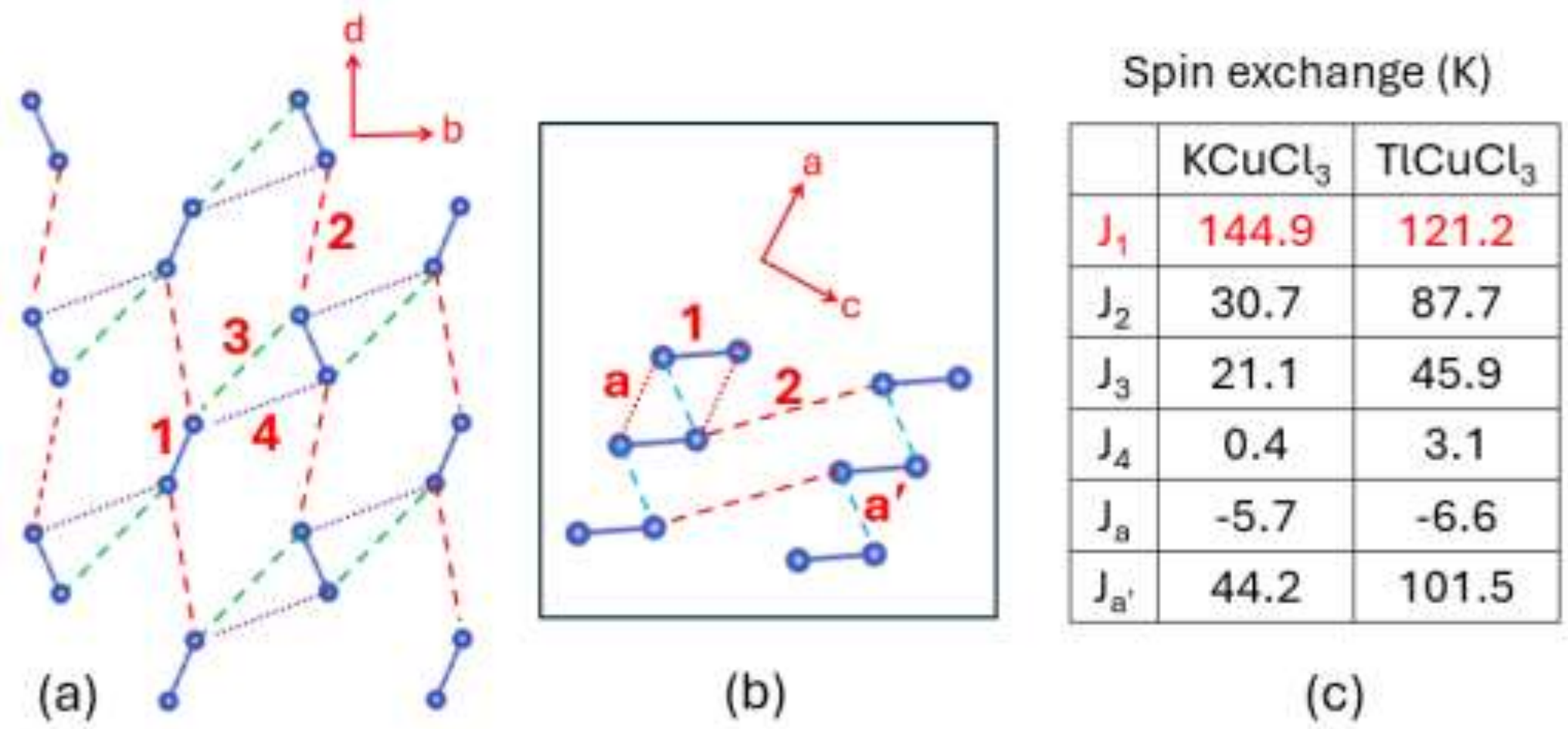

| | $KCuCl_3$ | $TlCuCl_3$ |
|---|---|---|
| $J_1$ | 144.9 | 121.2 |
| $J_2$ | 30.7 | 87.7 |
| $J_3$ | 21.1 | 45.9 |
| $J_4$ | 0.4 | 3.1 |
| $J_a$ | -5.7 | -6.6 |
| $J_{a'}$ | 44.2 | 101.5 |

**Fig. 5**. Intra- and interdimer spin exchanges of $KCuCl_3$ and $TlCuCl_3$: (a) Spin exchanges between $Cu_2Cl_6^{2-}$ spin dimers in each layer parallel to the bd-plane (i.e., the cleavage plane), where **d** = **a** + **c**/2 in (a), and those parallel to the ac-plane in (b). Note that the bd-plane is the cleavage plane. Only the Cu atoms of each $Cu_2Cl_6^{2-}$ anion are shown for simplicity. The labels 1 – 4, a and a′ refer to the spin exchanges $J_1$ – $J_4$, $J_a$ and $J_{a'}$, respectively. (c) Spin exchanges of $KCuCl_3$ and $TlCuCl_3$ determined by DFT+U calculations.[1]

The intradimer bond $J_1$ (i.e., $\Delta$ to a first approximation) is larger for $KCuCl_3$ than for $TlCuCl_3$ (i.e., 145 vs. 121 K). Therefore, at 1.7 K under $\mu_0 H = 0$, the Boltzmann factor $\exp(-\Delta/k_B T)$ is greater for $TlCuCl_3$ than for $KCuCl_3$ by a factor of $1.34\times10^6$. As the field increases from 0 to $\mu_0 H_d$, the values of $\exp(-\Delta_H/k_B T) = \exp[(-\Delta/k_B T)(1 - H/H_d)]$ (**Fig. 3**) for $TlCuCl_3$ and $KCuCl_3$ decrease exponentially to 1. This implies that, at a given $\mu_0 H$ lower than $\mu_0 H_d$, $\delta E_Z$ is much greater for $TlCuCl_3$ than for $KCuCl_3$. Then, the magnetic field $\mu_0 H_c$ needed to start breaking the intradimer magnetic bonds $J_1$ will be much lower for $TlCuCl_3$. This explains why $TlCuCl_3$ has a much narrower zero-magnetization plateau than does $KCuCl_3$ (**Fig. 4c**) despite that the interdimer spin exchanges are much stronger for $TlCuCl_3$ (**Fig. 5c**).

Matsumoto et el.[11] fitted the magnon dispersion relations (i.e., the energies of their magnetic states as a function of the repeat wave vectors) of $TlCuCl_3$ and $KCuCl_3$, resulting from the inelastic neutron scattering measurements carried out for single crystal samples by Cavadini et al.,[12,13] in terms of five fitting parameters (one intradimer and three interdimer spin exchanges) and one energy gap (which is indirectly related to spin exchanges). Their fitting analysis led to the result that $J_1$ is greater for $TlCuCl_3$ than for $KCuCl_3$ (i.e., of 64 vs. 49 K). In addition, their strongest interdimer bond, $J_2$, is stronger for $TlCuCl_3$ than for $KCuCl_3$ (i.e., 37 vs. 9 K). Thus, with the intra- and interdimer spin exchanges extracted by Matsumoto et el.,[11] one cannot explain the experimental observation that the width of the zero-magnetization plateau is much narrower for $TlCuCl_3$ than for $KCuCl_3$.

### 3.2.C. Anisotropy in the zero-magnetization plateau of $KCuCl_3$

The zero-magnetization width of $KCuCl_3$ is slightly wider for the $H$||cp than for the $H$⊥cp field (i.e., ~22 vs ~20 T).[5] This observation is related to the fact that the g-factor of the $Cu^{2+}$ ion at a square planar site is anisotropic, and that $g_{||z} = 2.22$ and $g_{\perp z} = 2.05$,[16] so $g_{\perp cp} \approx 2.22$ and $g_{||cp} \approx 2.05$. This makes the spin moments of the $Cu^{2+}$ ion anisotropic, i.e.,

$$|\mu_s(\perp\mathrm{cp})| \approx \mu_B g_{\perp\mathrm{cp}} S > |\mu_s(||\mathrm{cp})| \approx \mu_B g_{||\mathrm{cp}} S. \qquad (4)$$

At a given external field, therefore, the Zeeman energy is greater for $H$⊥cp than for $H$||cp. Consequently, the field needed to break all intradimer bonds will be lower for $H$⊥cp than for $H$||cp , namely, $H_c(\perp\mathrm{cp}) < H_c(||\mathrm{cp})$, in agreement with experiment.[5] Note that the ⊥cp direction deviates slightly from the ||z direction (**Fig. 4b**), so it is of interest to examine a possible effect of this deviation on our analysis of the above anisotropy. Given the difference $\Delta g = g_{||z} - g_{\perp z} = 2.22 – 2.05 = 0.17$, and θ as the angle of the deviation from the z-axis, the dependence of the g-factor on the deviation angle can be approximated by $g(\theta) = 2.22 – \Delta g \sin\theta$. Then, $g(5°) = 2.20$ and $g(10°) = 2.19$. These values are quite close to $g_{||z}$. Thus, the anisotropy argument presented above holds good.

### 3.3. Specific heat behavior of an antiferromagnet with type II zero-magnetization plateau

This section analyzes how external magnetic field affects the magnetic entropy of an antiferromagnet of type II zero-magnetization plateau and shows that this prediction can be verified by measuring its specific heat as a function of applied magnetic field.

#### 3.3.A. Magnetic entropy

We first examine the magnetic entropy associated with type II zero-magnetization plateaus and its implication. As discussed for an AFM chain of AFM dimers, the number of broken interdimer bonds increases with increasing the number of spin-flipped AFM dimers in one of the two sets of AFM dimers (**Fig. 2b** and **2c**). According to Le Chatelier's principle, the number of spin-flipped AFM dimers in one set of the AFM dimers would increase. Suppose that the AFM chain consists of *n* AFM dimers. Then, each of the two sets of AFM dimers has *n*/2 dimers. At a given field below $\mu_0 H_m$ (**Fig. 2g**), there is a certain number (say, *m*) of spin-flipped AFM dimers in one set of the dimers. Since these spin-flipped dimers can be anywhere within the set of the dimers, there are Ω(*m*) ways of distributing *m* dimers in the *n*/2 sites of the set, where Ω(*m*) is the binomial coefficient, i.e.,

$$\Omega(m) = {}_{n/2}\mathrm{C}_{m} \tag{5}$$

Thus, the associated configurational magnetic entropy is given by

$$\mathrm{S}(m) = k_B ln\Omega(m). \tag{6}$$

Note that Ω(*m*) is a symmetric function with maximum at *m* = *n*/4 and minimum (i.e., 1) at *m* = 1 and *n*/2, and hence so is S(*m*). In the field between $\mu_0 H_m$ and $\mu_0 H_c$, the spin arrangement of the AFM chain is described by one spin arrangement (i.e., the one given in **Fig. 2d**) leading to zero magnetic entropy.

### 3.3.B. Magnetic entropy from specific heat measurements

To verify the above conclusion concerning a type II magnetization plateau, it is necessary to measure the field-dependent specific heat, *C*(*H*), at a low temperature in the field region covering the zero magnetic plateau, as in the study of γ-$Mn_3(PO_4)_2$.[2] The $C(H)$ has two contributions, namely,

$$C(H) = C_{ph}(H) + C_m(H), \quad (7)$$

where $C_{ph}(H)$ and $C_m(H)$ are the phonon and magnetic entropy contributions, respectively.[2] The internal energy $E_{ie}(H)$ of an antiferromagnet is expected to increase with field $\mu_0H$ by a series of events governed by Le Chatelier's principle;[2] increasing Zeeman energy enhances the spin-lattice interactions, which in turn increase the vibrational energy, eventually raising the internal energy $E_{ie}(H)$ and hence lowering $C_{ph}(H)$. According to the specific heat study of γ-$Mn_3(PO_4)_2$, it is a good approximation that $E_{ie}(H)$ increases linearly with field, so $C_{ph}(H)$ decreases linearly with field. We will assume this approximation to be also valid for $KCuCl_3$. It is commonly believed that, in most solids, the phonon contribution is essentially field-independent, and any field dependence due to magnetostriction is extremely small compared to the magnetic contribution. However, recent experimental studies[17-20] showed that this belief is unjustified under high magnetic field.

In the following analysis, our discussion is qualitative because the field-dependence of the phonon contribution to the internal energy and hence to the specific heat is unknown. Consequently, it is not possible to quantitatively simulate the field-dependence of the specific heat experimentally observed for $KCuCl_3$ (see below) in terms of the specific heat predicted from the magnetic entropy discussed above. Thus, our discussion in the following is aimed at establishing that the field dependence of the specific heat observed for $KCuCl_3$ is fully consistent with the prediction from the field-dependence of the magnetic entropy.

In the previous section, we noted that the magnetic entropy is described by $k_B ln\Omega(m)$ in the field region between 0 and $\mu_0H_m$, but it is 0 in the field region between $\mu_0H_m$ and $\mu_0H_c$ (**Fig. 2g**).

Given $\mu_0 H_{max}$ as the field where $C_m(H)$ reaches its maximum, i.e., $C_m(H_{max})$, the phonon contribution at $\mu_0 H_{max}$ is written as $C_{ph}(H_{max})$. The $C(H)$ vs. $H$ curve expected for the case when $C_m(H_{max}) > C_{ph}(H_{max})$ is schematically shown in **Fig. 6a**, where $C(H) = C_{ph}(H) + C_m(H)$ in the $0 - \mu_0 H_m$ region, but $C(H) = C_{ph}(H)$ in the $\mu_0 H_m - \mu_0 H_c$ region. In addition, $C_{ph}(H)$ is assumed to decrease linearly with field because this approximation was found to well describe the $C(H)$ of γ-$Mn_3(PO_4)_2$.[2] The $C(H)$ peak of **Fig. 6a** is narrower than the $C_m(H)$ peak because both the left-hand and the right-hand sides of the $C_m(H)$ peak are reduced by $C_{ph}(H)$. Furthermore, the reduction is stronger on the right- than on the left-hand side of the peak. Thus, for an antiferromagnet exhibiting a zero-magnetic plateau in the $0 - \mu_0 H_c$ region, the above discussion predicts that:

(a) The specific heat $C(H)$ has a broad peak in the $0 - \mu_0 H_m$ region because of the magnetic entropy $C_m(H)$.

(b) This broad peak becomes asymmetric due to the phonon contribution $C_{ph}(H)$, leading to the result

$$C(0) > C(H_m) \qquad (8)$$

(c) In the $0 - \mu_0 H_m$ region, the broad $C(H)$ peak is wider on the left- than on the right-hand side.

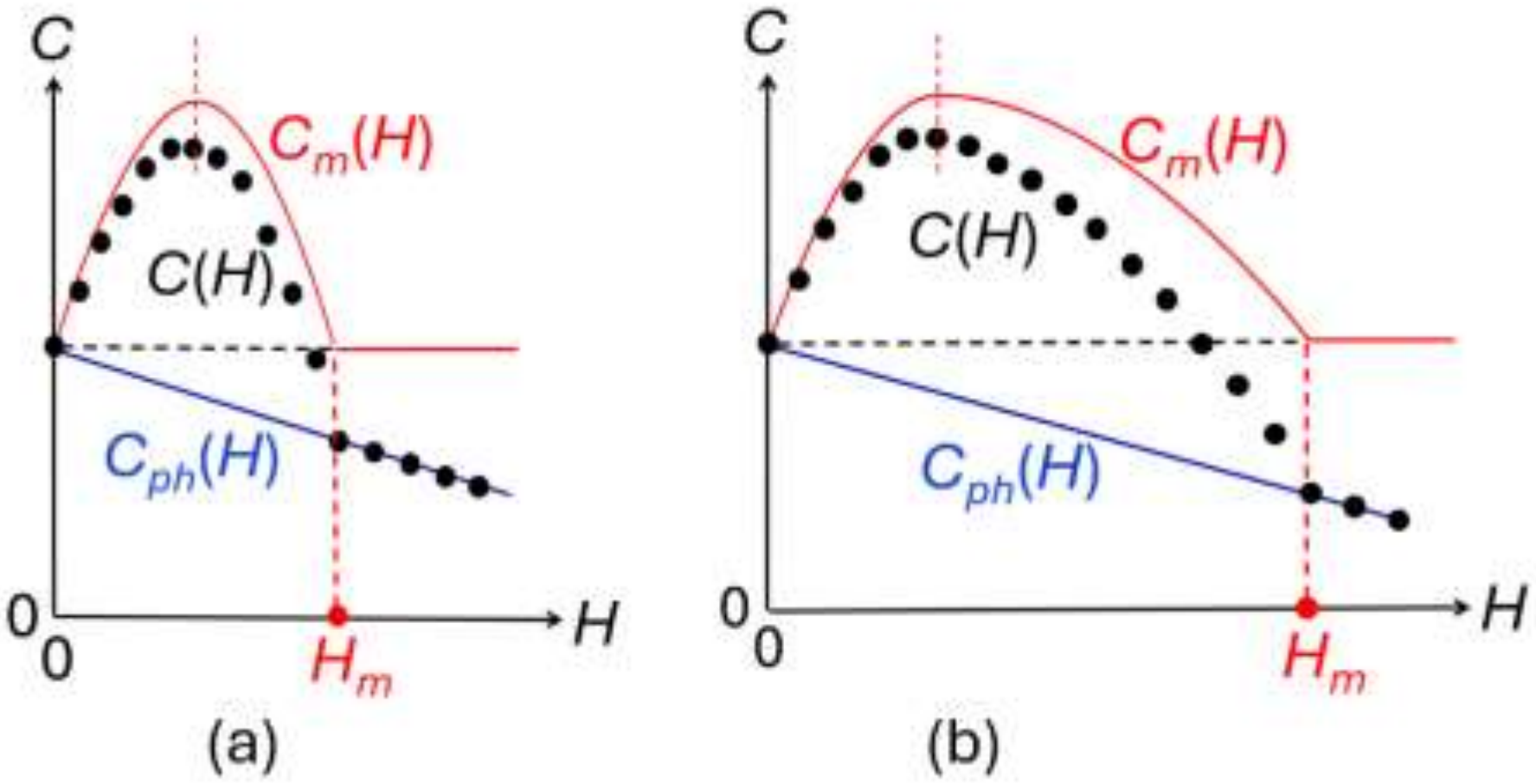

**Fig. 6**. Schematic diagrams depicting the field-dependence of the specific heat $C(H)$ expected for an antiferromagnet exhibiting a zero-magnetization plateau in the field region between 0 and $\mu_0 H_m$ for cases when $C_{ph}(H)$ decreases linearly with $H$. (a) Case of a symmetric $C_m(H)$ when the left- and right-hand sides of the peak are the mirror images of each other. (b) Case of an asymmetric $C_m(H)$ when the left-hand side is considerably narrower than the right-hand side of the peak. $C_{ph}(H)$ is represented by a solid purple curve, $C_m(H)$ by a solid red curve, and $C(H)$ by a black-dot curve. The vertical dashed line in (a) and (b) represents the position of $C_m(H_{max})$.

### 3.3.C. Field-dependent specific heat measurement for $KCuCl_3$

In this section we verify the conclusions of the previous section by studying how the specific heat of $KCuCl_3$ is affected by external magnetic field. The $\mu_0 H_c$ value of $KCuCl_3$ (greater than ~20 T) is well beyond the highest field (i.e., 9 T) available from standard laboratory equipment. Nevertheless, $KCuCl_3$ is still an interesting system because specific heat measurements will allow us to test the three predictions of the previous section concerning the $C(H)$ behavior in the 0 – $\mu_0 H_m$ region.

The temperature-dependent specific heat $C_p(T)$ of $KCuCl_3$ were measured at 0 T and 9 T (**Fig. S2**) and the resulting data were analyzed using the Debye and two Einstein functions (**Section S2** in the supporting information). When the magnetic contribution to the specific heat is fitted to the model of a spin-half AFM dimer, we obtain the spin gap $\Delta$ of 54 K at 0 T and 51 K at 9 T. The reduction of the gap under magnetic field is expected, because the spin gap under field $\Delta_H$ should become smaller than $\Delta$ due to the field-induced split of the triplet state (**Fig. 3**).

Results of our field-dependent specific heat measurements for $KCuCl_3$ at 2 K are presented in **Fig. 7**, which shows that the $C_p(H)$ vs. $\mu_0 H$ plots obtained while increasing and decreasing the magnetic field coincide completely. The $C_p(H)$ has a broad peak, which is asymmetrical with respect to the $\mu_0 H_{max}$ where the maximum of the $C_p(H)$ peak occurs. Furthermore, $C_p$(0 T) > $C_p$(9 T). These observations confirm the two predictions, (a) and (b), of the previous section. However, the observed $C_p(H)$ peak is much wider on the right- than on the left-hand side of the peak, which is opposite to the prediction (c).

In arriving at the prediction (c), it was implicitly assumed that the magnetic entropy S($m$) leading to the magnetic specific heat $C_m(H)$ is solely determined by the binomial coefficients, namely, by the statistical probability. Let the constant $\mu_0 \delta H_0$ be the field required for the antiferromagnet in a two-phase region with $m$ spin-flipped spin dimers to add one additional spin-flipped spin dimer. Since $\mu_0 \delta H_0$ is independent of $m$, the increase of $m$ from 1 to $n/2$ corresponds to the field increase from 0 to $\mu_0(n\delta H_0/2) = \mu_0 H_m$. If this is the case, the $C_m(H)$ vs. $H$ plot would have a broad symmetrical peak as depicted in **Fig. 6a**.

However, one must consider how the energy required for the spin-flip of an additional spin dimer depends on the number of spin-flipped spin dimers already present in the antiferromagnet.

Each spin-flipping brakes two interdimer bonds, so the antiferromagnet becomes steadily more unstable as it increases the number of spin-flipped spin dimers. This implies that it requires more energy to spin-flip an additional spin dimer, as the antiferromagnet has more spin-flipped spin dimers. Then, the field $\delta H(m)$ required for the antiferromagnet with $m$ spin-flipped spin dimers to have an additional spin-flipped spin dimer should increase with $m$. The simplest form for $\delta H(m)$ can be written as

$$\delta H(m) = [1 + (m - 1)\alpha]\delta H_0, \qquad (9)$$

where $\alpha$ is a positive constant. Thus, the interval $\delta H(m)$ between adjacent $m$ points increases with $m$. That is, on the magnetic field axis, the distance between adjacent points, corresponding to the $m$ points, becomes larger with increasing $m$. This has the effect of stretching out the symmetrical $C_m(H)$ vs. $H$ curve (**Fig. 6a**) toward the higher field direction such that the stretching out becomes progressively stronger with $m$. Thus, the resulting $C_m(H)$ vs. $H$ becomes asymmetrical as depicted in **Fig. 6b**, widening the right-hand side more than the left-hand side of the broad peak, as depicted in **Fig. 6b**. Then, in the resulting $C(H)$ vs. $H$ plot, the broad peak will be wider on the right- than on the left-hand side of the peak, as found experimentally (**Fig. 7**).

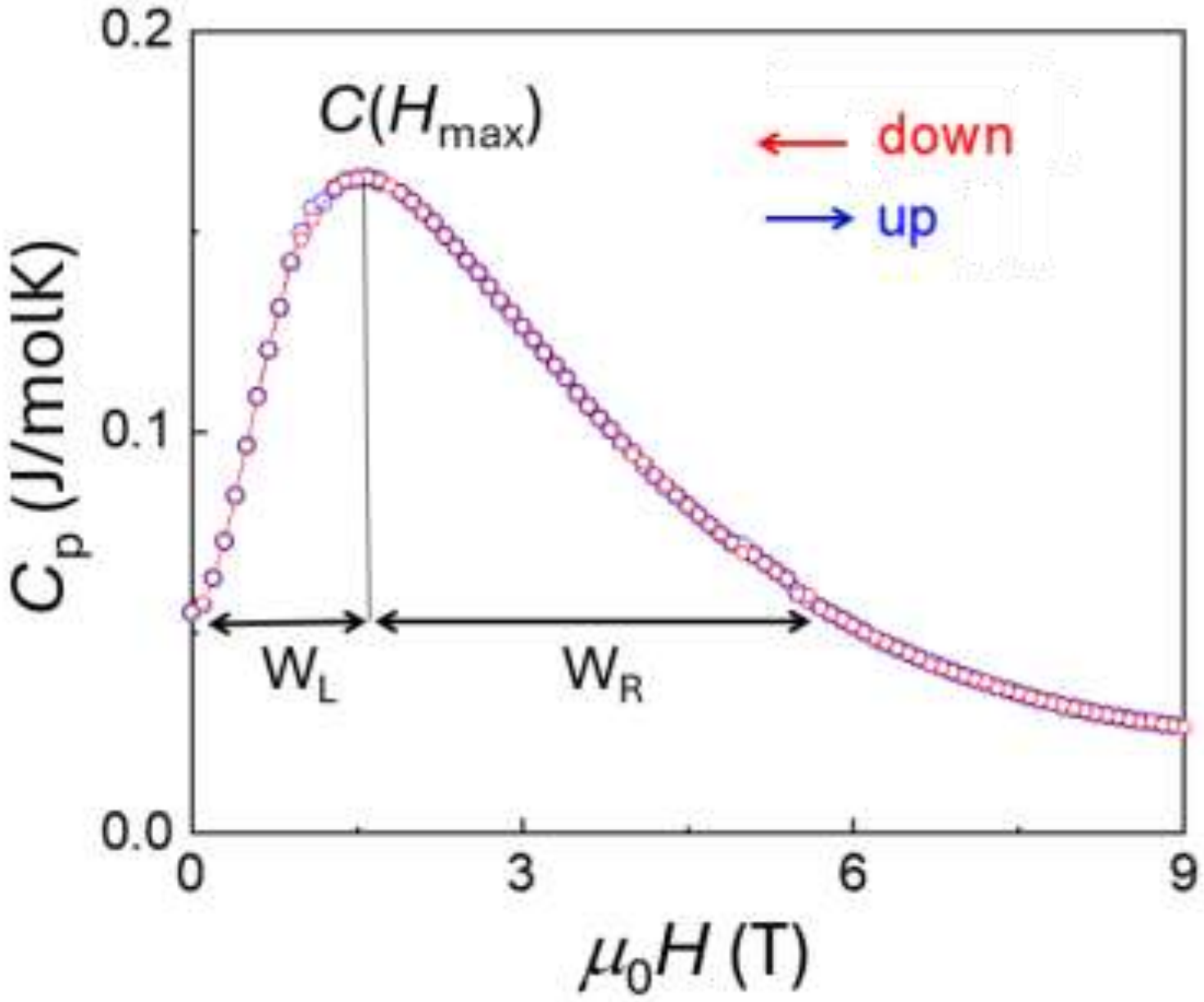

**Fig. 7**. Field dependence of the specific heat of $KCuCl_3$ measured at 2 K.

**4. Discussion**

In describing the magnetic entropy for the two-phase region of a nonzero-magnetization plateau, it was found that the binomial distribution of partitioned-out ferrimagnetic fragments in the spin lattice of $\gamma$-$Mn_3(PO_4)_2$ [2] is determined by the modified binomial distribution $\langle\Omega(m)\rangle$. This arises from the way the ferrimagnetic layers are antiferromagnetically coupled leading to a one-dimensional character. With increasing the number of partitioned-out layers, partitioning out one more can be accompanied by an additional one.[2] In an ideal trigonal antiferromagnet, $RbFe(MoO_4)_2$,[1,21] which exhibits a 1/3-magnetization plateau, partitioning-out one more triangular unit from its two-dimensional triangular spin lattice would be accompanied by many more additional ones than found for $\gamma$-$Mn_3(PO_4)_2$,[2] leading to a more strongly modified $\langle\Omega(m)\rangle$. In the two-phase region of zero-magnetization plateaus of $KCuCl_3$ and $TlCuCl_3$, we found that the magnetic entropy $S(m) = k_B ln\Omega(m)$ for distributing $m$ spin-flipped spin dimers in the spin lattice is given by the binomial distribution $\Omega(m)$. However, in increasing the number $m$ by 1, the required field is not constant but increases with $m$. How to quantitatively describe the modified binomial distribution $\langle\Omega(m)\rangle$, details of which would depend on the nature of magnets under study, is an important issue to study in the future.

It is of interest to ask whether all the interdimer bonds of $KCuCl_3$ are broken within 9 T, namely, if the field $\mu_0 H_m$ lies within 9 T. Although the specific heat is lower at 9 T than at 0 T (**Fig. 7**), the $C_p(H)$ value can be considerably lower at $\mu_0 H_m$ than at 0 T according to **Fig. 6a** and **6b**. Thus, it is not certain if $\mu_0 H_m < 9$ T in $KCuCl_3$. Specific heat measurements under a wider range of magnetic fields are necessary to resolve this issue. Before leaving this section, we point out that $TlCuCl_3$ is an excellent system to study because its zero-magnetization plateau region (0 – ~6 T) as well as the nonzero-slope region immediately following the zero-magnetization plateau lies in the easily accessible field regime (0 – 9 T) of standard laboratory equipment. However, it is difficult to prepare $TlCuCl_3$ due to the safety concern related to the highly toxic nature of Tl and its compounds needed for the synthesis of $TlCuCl_3$.

## 5. Concluding remarks

In antiferromagnets of antiferromagnetically coupled AFM dimers under magnetic field, the occurrence of a zero-magnetization process is driven by the field-induced local excitations in each spin dimer, which leads the excited ($S = 1$) state to mix into the ground state ($S = 0$) in each AFM spin dimer. The mixing of the excited state by the amount of the Boltzmann factor, $\exp(-\Delta/k_B T)$, provides each AFM spin dimer with a minute nonzero spin moment, hence enabling the antiferromagnet to absorb Zeeman energy. This explains the experimental observations that the width of the zero-magnetization plateau is much narrower for $TlCuCl_3$ than for $KCuCl_3$ (~6 vs. ~22 T) as well as the anisotropy of the zero-magnetization plateau found for $KCuCl_3$. The occurrence of zero-magnetization plateaus in antiferromagnets of AFM spin dimers provides another example in support of the concept that antiferromagnets under magnetic field can absorb Zeeman energy by field-induced local excitations.

From the viewpoint of configurational magnetic entropy, the field region of a zero-magnetization plateau is subdivided into two regions, i.e., the $0 – \mu_0H_m$ region (a two-phase region), where AFM dimers with broken interdimer bonds and those with no broken interdimer bonds exist together, and the $\mu_0H_m – \mu_0H_c$ region (a single-phase region), where only the AFM dimers with broken interdimer bonds exist. The magnetic entropy of the $0 – \mu_0H_m$ region is described by $k_B ln\Omega(m)$ while that of the $\mu_0H_m – \mu_0H_c$ region is 0. The specific heat $C_p(H)$ measured at 2 K for $KCuCl_3$ between 0 and 9 T exhibits a broad asymmetric peak clearly reflecting the contribution of the magnetic entropy $C_m(H)$ expected for the $0 – \mu_0H_m$ region.

**Acknowledgements**

OSV thanks the support from RSCF grant No. 25-12-00028 for specific heat measurements. The research at KHU was supported by Basic Science Research Program through the National Research Foundation of Korea (NRF) funded by the Ministry of Education (RS-2020-NR049601).

**Conflicts of interest**

There are no conflicts to declare.

**Data availability**

The data supporting this article has been included as part of the Supplementary Information.

**Supplementary material**

Supplementary Section S1 on the spin orientation of $KCuCl_3$ and $TlCuCl_3$ with Fig. S1, as well as Section S2 on the temperature dependence of the specific heat measured for $KCuCl_3$ at 0 and 9 T with Fig. S2. All are in pdf.

TOC figure

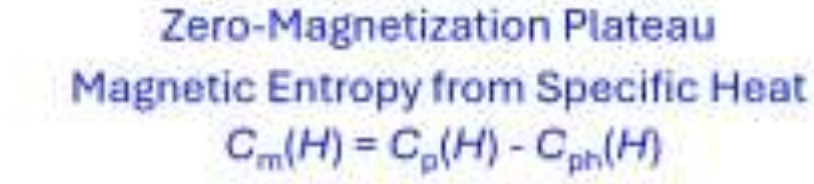
Zero-Magnetization Plateau
Magnetic Entropy from Specific Heat
$C_m(H) = C_p(H) - C_{ph}(H)$

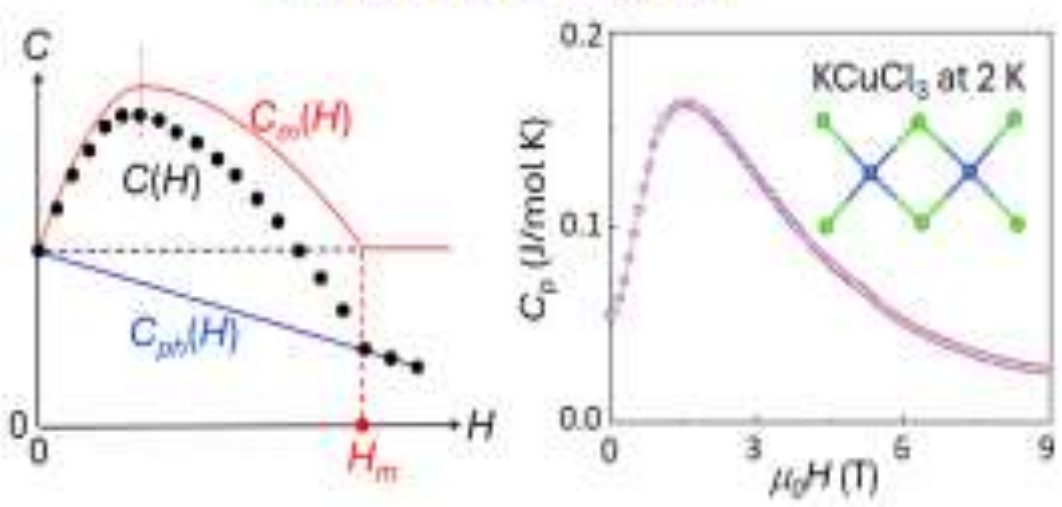
C
$C_m(H)$
$C(H)$
$C_{ph}(H)$
0
0
H
$H_m$
0.2
0.1
0.0
$C_p$ (J/mol K)
KCuCl$_3$ at 2 K
0
3
6
9
$\mu_0 H$ (T)

Supporting Information

for

**Field-Induced Local Excitations Causing Zero-Magnetization Plateaus in Antiferromagnets of Antiferromagnetic Spin Dimers Under Magnetic Field**

Myung-Hwan Whangbo[1,*], Hyun-Joo Koo[2], Nikita V. Astakhov,[3,4] Peter S. Berdonosov,[4] Olga S. Volkova[5,*]

[1] Department of Chemistry, North Carolina State University, Raleigh, NC 27695-8204, USA

[2] Department of Chemistry, Research Institute for Basic Sciences, Kyung Hee University, Seoul 02447, Republic of Korea

[3]Faculty of Materials Sciences, Lomonosov Moscow State University, Moscow 119991, Russia

[4]Faculty of Chemistry, Lomonosov Moscow State University, Moscow 119991, Russia

[5]Faculty of Physics, Lomonosov Moscow State University, Moscow 119991, Russia

mike_whangbo@ncsu.edu

os.volkova@yahoo.com

## S1. Spin orientation of $KCuCl_3$ and $TlCuCl_3$

A transition-metal magnetic ion M forms a $ML_n$ (typically, n = 3 – 6) polyhedron with its surrounding main group ligands L. We take the rotational axis of this polyhedron as the z-axis. The preferred spin orientation of M is predicted by the selections rule[1] based on the interaction of the highest occupied d-state with the lowest unoccupied d-state of the $ML_n$. This interaction is induced by the spin-orbit coupling (SOC) of the magnetic ion M, which leads to the selection rule based on the minimum difference, $|\Delta L_z|$, in the magnetic quantum numbers of these two states:

If $|\Delta L_z| = 0$, the ||z orientation is predicted.

If $|\Delta L_z| = 1$, the $\perp$z orientation is predicted.

If $|\Delta L_z| > 1$, there is no SOC-induced interaction.
In such a case, the interaction of the next highest occupied state with the lowest unoccupied state will decide the preferred spin orientation.

With the local Cartesian coordinate given for a square planar $CuCl_4$ unit as in **Fig. S1a**, the d-states of $CuCl_4$ are split as depicted in **Fig. S1b**. In the spin-polarized description of the d-sates for a $Cu^{2+}$ ($d^9$) ion, the up-spin states lie below the down-spin states, so the highest occupied and lowest unoccupied states in the down-spin states as shown in **Fig. 1Sb**. The $z^2$ state does not interact with the $x^2$-$y^2$ state because their $|\Delta L_z| = 2$. Thus, one needs to consider the interaction of the xy state with the $x^2$-$y^2$ state. Since their $|\Delta L_z| = 0$, the preferred spin orientation is the ||z direction. The observed spin orientations of the $Cu^{2+}$ ions in $KCuCl_3$[2] and $TlCuCl_3$[3] are approximately the $\perp$z direction. The latter is explained if the split d-states of their $Cu_2Cl_6^{2-}$ ion (**Fig. S1c**) has the xy state lying lower than the (xz, yz) state as depicted in **Fig. S1d**, because the interaction of the (xz, yz) state with the xy leads to $|\Delta L_z| = 1$. The lowering of the xy state below the (xz, yz) state occurs because the two xy orbitals make a through-space bonding interaction across the shared edge of the two $CuCl_4$ units (**Fig. S1e**).

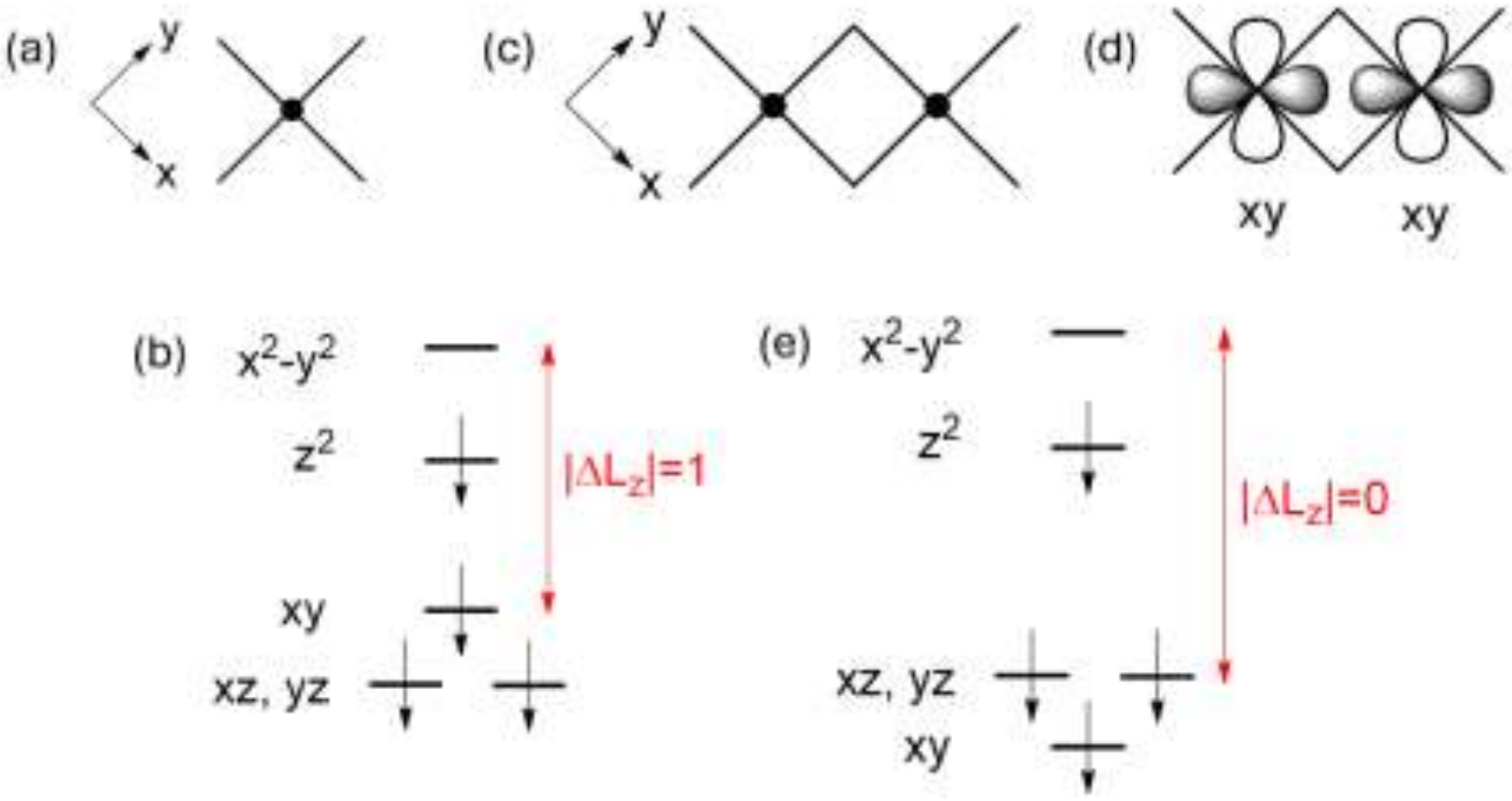

Fig. S1. (a) $CuCl_4$ with the local x- and y-axes chosen along the Cu-Cl bonds. (b) Split d-states expected for a $Cu^{2+}$ ion in $CuCl_4$. (c) $Cu_2Cl_6^{2-}$ ion with the local x- and y-axes chosen along the Cu-Cl bonds. (d) Split d-states expected for a $Cu^{2+}$ ion in the $Cu_2Cl_6^{2-}$ ion. (e) Bonding interaction between the xy orbitals of the two $Cu^{2+}$ ions in the $Cu_2Cl_6^{2-}$ ion.

**S2. Temperature-dependent specific heat of $KCuCl_3$**

To check the lattice contribution, $C_{lattice}$, the specific heats were measured at 0 T and 9 T as shown in **Fig. S2**. The lattice contribution was approximated using the Debye and two Einstein functions with weights $\Theta_D$ = 3.2 for $\alpha_D$ = 210 K, $\alpha_{E1}$ = 1.1 for $\Theta_{E1}$ = 790 K, and $\alpha_{E2}$ = 0.7 for $\Theta_{E2}$ = 480 K. The sum of $\alpha_D + \alpha_{E1} + \alpha_{E2} = 5$, which is equal to the number of atoms per formula unit. The magnetic entropy obtained after subtracting $C_{lattice}$ from $C_p$ yields 5.6 J/mol K (the upper inset of **Fig. S2**). This value is close to $S_m$ = Rln2 = 5.8 J/mol K. The fit of the magnetic contribution to the specific heat for a spin-half antiferromagnetic dimer,

$$C = \frac{3}{2}R\left(\frac{\Delta}{kT}\right)^2 \frac{\exp\left(-\frac{\Delta}{kT}\right)}{\left(1+3\exp\left(-\frac{\Delta}{kT}\right)\right)^2}, \qquad \text{(S1)}$$

leads to $\Delta$ = 54 K. The latter agrees with the value reported by Goto et al.[4] The lattice contribution $C_{lattice}$ from the $C_p(T)$ obtained at 9 T gives the broad maximum in $C_m(T)$ with $\Delta$ = 51 K, which agrees with a decrease in the AFM dimer gap under external magnetic field.

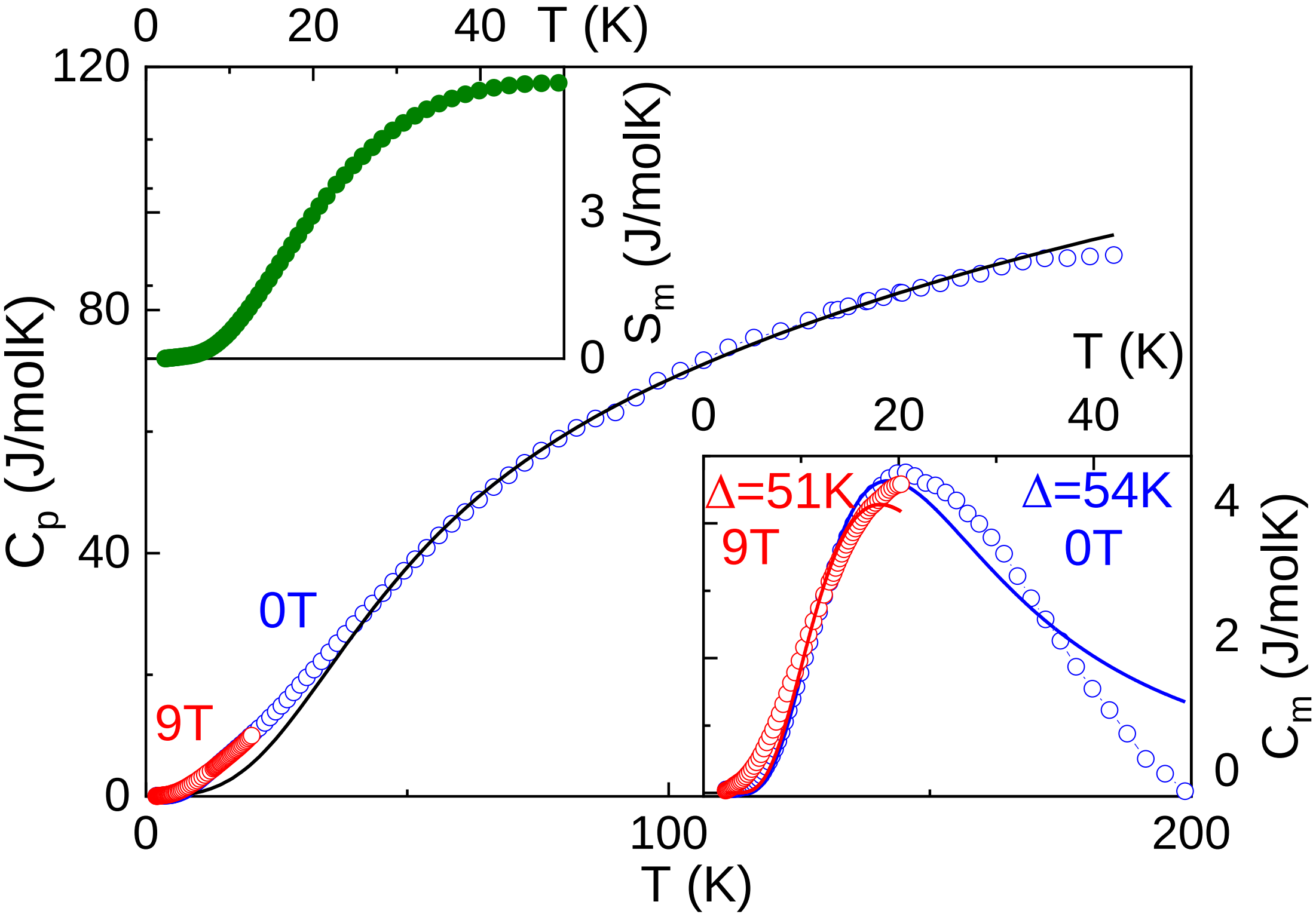

Fig. S2. Effect of field on the temperature dependent specific heat of $KCuCl_3$.